\documentclass[letterpaper, 10 pt, conference]{ieeeconf}  

\IEEEoverridecommandlockouts                              

\usepackage{graphics} 
\usepackage{epsfig} 

\usepackage{mathptmx} 
\usepackage{times} 
\usepackage{amsmath} 
\usepackage{amssymb}  
\usepackage{subcaption}
\usepackage{comment}
\usepackage{soul} 
\usepackage{cases} 
\usepackage[noadjust]{cite}
\usepackage{multicol}
\usepackage[dvipsnames]{xcolor}
\usepackage[font=small]{caption}

\newcommand{\revealing}{explicit}
\newcommand{\ambiguous}{ambiguous}
\newcommand{\numa}{n^a}
\newcommand{\numz}{z_{k}}
\newcommand{\numr}[1]{n^e_{#1}}
\newcommand{\cp}[1]{c_{#1}} 
\newcommand{\cm}[1]{\tilde{c}_{#1}} 
\newcommand{\deld}[1]{\delta d_{#1}}
\newcommand{\diffcp}[1]{\Delta c_{#1}} %
\newcommand{\diffcm}[1]{\Delta \tilde{c}_{#1}} %

\renewcommand{\baselinestretch}{.99}

\renewcommand{\b}{\mathbf{b}}
\newcommand{\e}{\mathbf{a}^E}

\newcommand{\vea}{\e} 

\usepackage{amsthm}
\newtheorem{theorem}{Theorem}
\newtheorem{lem}{Lemma}
\newtheorem{definition}{Definition}
\newtheorem{cor}{Corollary}

\newtheorem{rem}{Remark}

\newcommand{\game}[1]{\mathcal{G}_{#1}}

\newcommand{\figref}[1]{Fig.~\ref{#1}}

\newcommand{\edit}[1]{\textcolor{blue}{#1}}

\title{\LARGE \bf
The Eater and the Mover Game
}

\author{
Violetta Rostobaya$^1$, Yue Guan$^2$, James Berneburg$^1$, Michael Dorothy$^3$, and Daigo Shishika$^1$
\thanks{We gratefully acknowledge the support of ARL grant ARL DCIST CRA W911NF-17-2-0181. The views expressed in this paper are those of the authors and do not reflect the official policy or position of the United States Army, Department of Defense, or the United States Government.}
\thanks{$^1$Violetta Rostobaya, James Berneburg and Daigo Shishika are with College of Engineering and Computing at George Mason University, Emails: {\tt\footnotesize $\{$vrostoba,jbernebu,dshishik$\}$@gmu.edu}}
\thanks{$^2$Yue Guan is with the School of Aerospace Engineering, Georgia Institute of Technology, Email: {\tt\footnotesize yguan44@gatech.edu}}
\thanks{$^3$Michael Dorothy is with DEVCOM Army Research Laboratory, Email: {\tt\footnotesize michael.r.dorothy.civ@army.mil}}
}

\begin{document}

\maketitle
\thispagestyle{empty}
\pagestyle{empty}

\begin{abstract}
This paper studies the idea of ``deception by motion'' through a two-player dynamic game played between a Mover who must retrieve resources at a goal location, and an Eater who can consume resources at two candidate goals. 
The Mover seeks to minimize the resource consumption at the true goal, and the Eater tries to maximize it.
While the Mover has the knowledge about the true goal, the Eater cannot differentiate between the two candidates.
Unlike existing works on deceptive motion control that measures the deceptiveness through the quality of inference made by a distant observer (an estimator), we incorporate their actions to directly measure the efficacy of deception through the outcome of the game.
An equilibrium concept is then proposed without the notion of an estimator.
We further identify a pair of equilibrium strategies and demonstrate that if the Eater optimizes for the worst-case scenario, hiding the intention (deception by ambiguity) is still effective, whereas trying to fake the true goal (deception by exaggeration) is not.
\end{abstract}

\vspace{-0.1in}
\section{Introduction}
%
In competitive games with asymmetric information, players can sometimes leverage deception to alter decision making of the opponent and get a higher payoff~\cite{Hespanha2000}. A player may perform deception by sensor jamming~\cite{Yavin1987} or by controlling shared information~\cite{Basar2018}. In this work we draw attention to deception via direct perception, when a player does not have a communication channel, but instead, tries to deceive the opponent by moving in a particular way.

A number of existing works consider deception in the context of motion control.
A common formulation is to consider an agent that tries to reach its goal without making it obvious to an observing agent who is trying to infer the true goal~\cite{Ornik2018DeceptionIO,Dragan2015DeceptiveRM,Topcu2022resal,Topcu2022supcontrol}.
Specifically, \cite{Topcu2022resal} formalized the notion of \emph{ambiguity} (hiding information about true goal), and \emph{exaggeration} (moving towards a decoy goal to send a false signal) as ways to measure deceptiveness. 
A common assumption made in the works above is that the deceived agent uses an estimator, and the deceiver leverages the knowledge of how this estimator works.
Furthermore, the deceived agent is often so naive that it is not aware of the possibility of being deceived.

Another common assumption is that the observer is not expected to commit any action during deceiver's motion. 
The works on goal recognition performed by passive observer fall into this category
\cite{ Ramrez2010ProbabilisticPR,Masters2017DeceptiveP,Masters2019CostBasedGR,Wayllace2016GoalRD,Kulkarni2018ResourceBS,Masters2018CostBasedGR, Pereira2017LandmarkBasedHF}.
In the sense of agents' decision making, all of those works are ``one-sided''.


In this paper we are interested in studying the possibility of deception without making the assumptions discussed above.
Specifically, we do not employ an estimator for the observing player, but instead, consider a payoff function that leads to a risk averse behavior.
We also require the observing agent to take an action at every time step, which is an important feature of our problem that allows us to measure the effectiveness of deception directly through the outcome of the game.

The contributions of this work are: 
(i) the formulation of a novel game that explores the effectiveness of deception against an observing agent that actively takes actions; 
(ii) the identification of the equilibrium strategies that do not use an estimator to predict opponent's behavior; and (iii) expression for the associated outcome of the game. 
Our results indicate that deception by exaggeration does \emph{not} work in our problem when the payoff function for the observer captures the worst-case scenario. 
Nevertheless, the deceiving agent can still use ambiguity to improve its payoff. 
This result is useful for us in investigating the settings for deception to exist. 

\begin{figure}[t]
\centering
  \begin{subfigure}{0.4\columnwidth}
  \includegraphics[width=\textwidth]{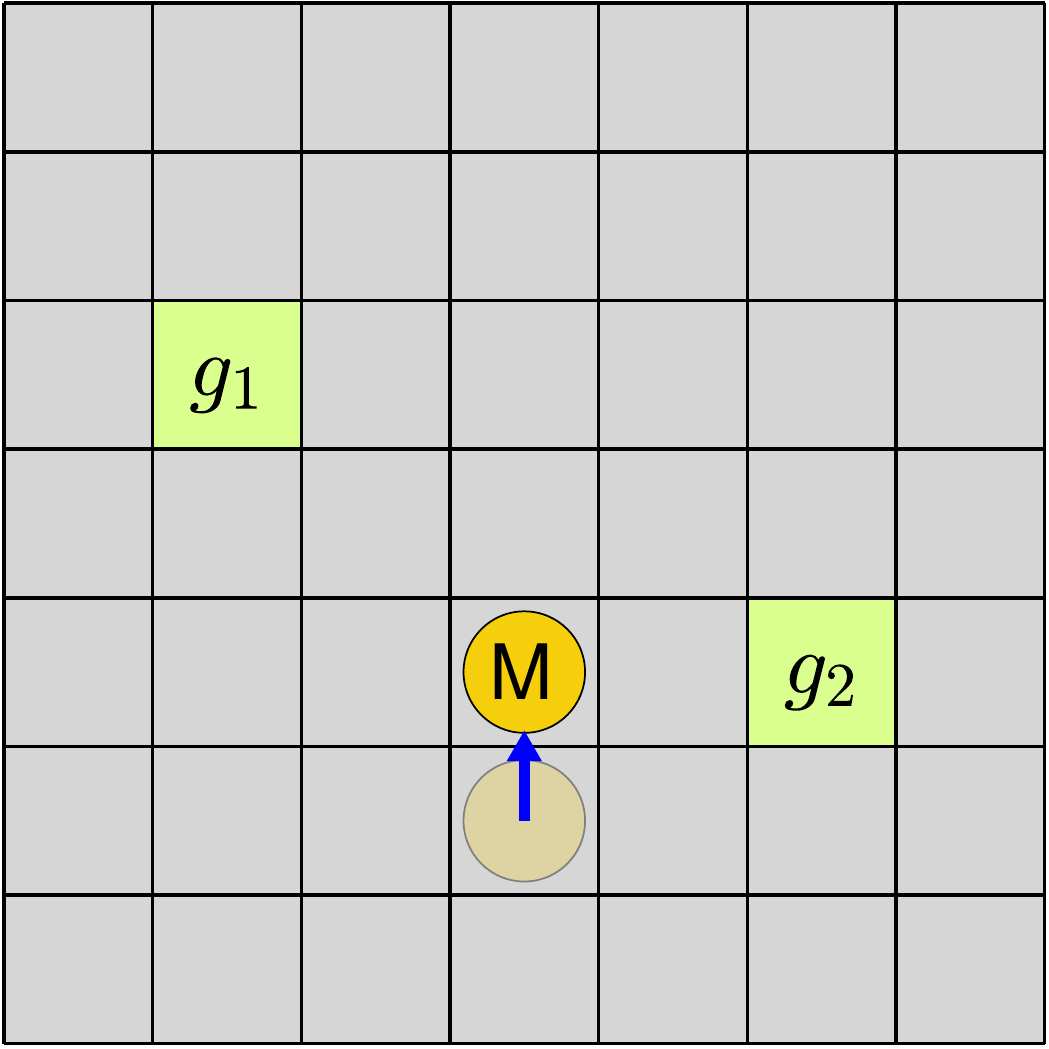}
  \caption{}\label{fig1a}
  \end{subfigure}
  \hspace{1em}
  \begin{subfigure}{0.4\columnwidth}
  \includegraphics[width=\textwidth]{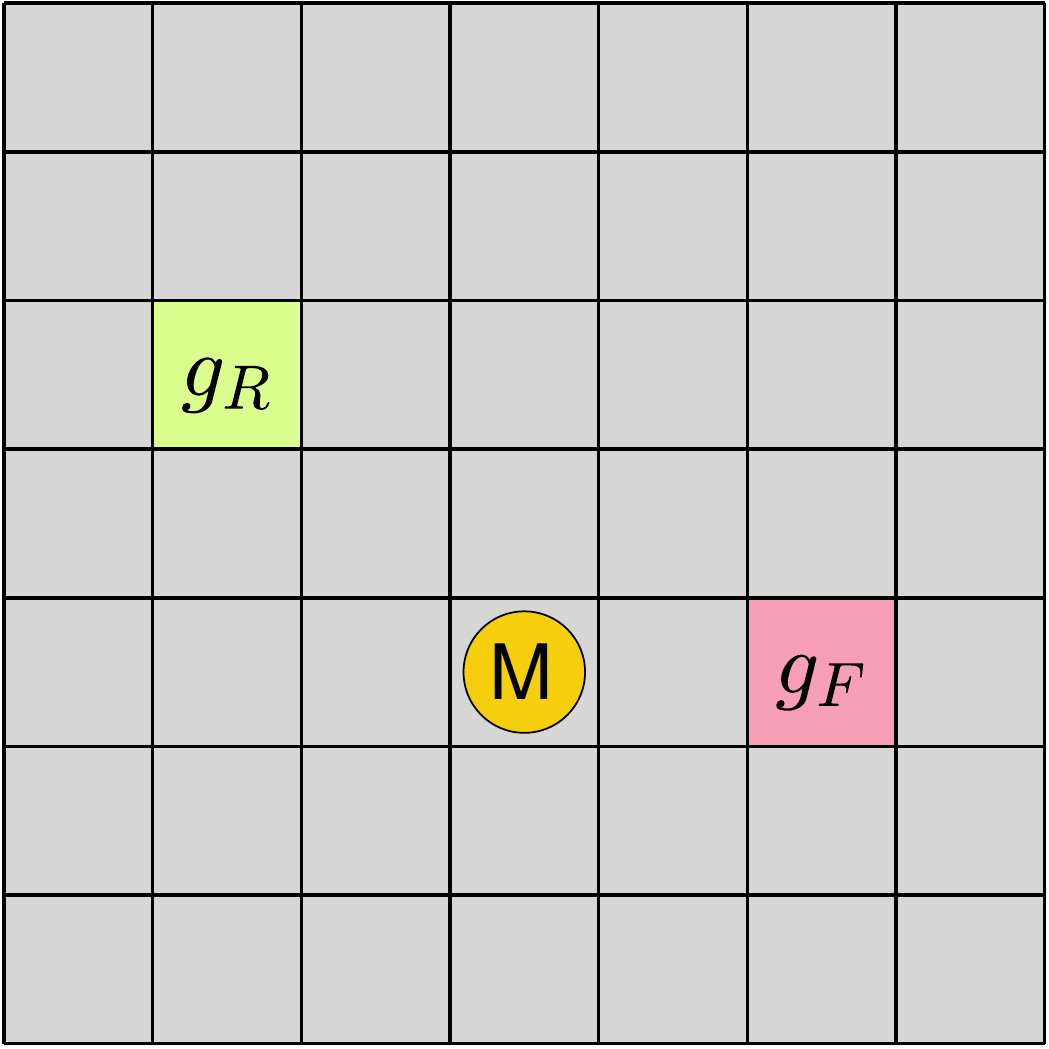}
  \caption{}\label{fig1b} 
  \end{subfigure} 
  \hfill
  \caption{Illustration of the asymmetric information. (\subref{fig1a}) Perspective of the Eater, who knows the most recent Mover's action (blue arrow). (\subref{fig1b}) Perspective of the Mover who knows the true goal $g_R$.}
  \label{fig:game}
  \vspace{-0.2in}
\end{figure} 
\section{Problem Formulation}
We consider a two-player discrete-time dynamic game with asymmetric information played between the Mover and the Eater in a grid world.
At the beginning of the game, two goal locations are specified by nature: $g_1,g_2 \in \mathbb{Z}^2$.
One is the true goal, $g_R$, and the other is the fake goal, $g_F$.
Each goal is given sufficiently large number of resources (bananas) at the beginning of the game.

\paragraph*{States}
The game evolves with two types of states.
One is the Mover's position $p(t)\in P$ where $P= \mathbb{Z}^2$. 
The other state is related to the number of bananas at each goal location, which the Eater  consumes.
We use $b_i(t) \geq 0$ to denote the consumption of bananas from $g_i$ for $i\in\{1,2\}$, and define $\b(t)=[b_1(t),b_2(t)]\in B$ as the \emph{consumption vector},\footnote{The difference between the numbers of bananas at the two goals at $t=0$ can be properly reflected by setting $b_i(0)>0$ for one of the goals.}
where $B = \mathbb{R}_{\geq 0}^2$ and $\b(0)\in\mathbb{Z}_{\geq 0}^2$.

\paragraph*{Actions}
The Mover selects an action $a^M(t)$ from its action space $A^M = \{ \texttt{up, down, left, right}\}$ at each time step, which updates its position from $p(t)$ to $p(t+1)$.
The Eater's action $\e(t)$ is drawn from $A^E = \{[1,0],[0,1], [0.5,0.5]\}$,
corresponding to eating one banana from $g_1$, eating one from $g_2$, or eating half from both goals.\footnote{We assume that the bananas at both goals do not run out.}
The consumption vector has the following dynamics:\footnote{This makes $b_i(t)$ to be a non-negative integer multiple of 0.5.}
\begin{equation*}
    \b(t+1) = \b(t) + \e(t).
\end{equation*}

\begin{figure}[t]
    \centering
    \includegraphics[width = 0.48\textwidth]{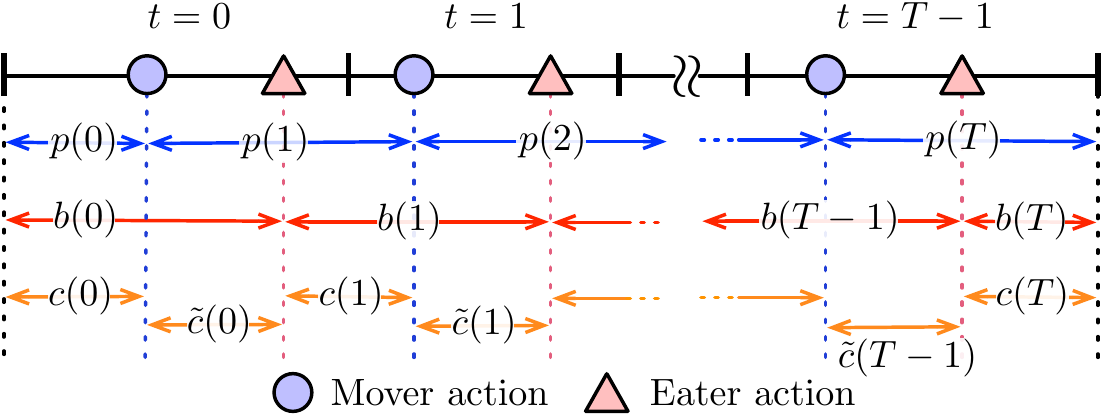}
    \caption{Illustration of the game timeline.}
    \label{fig:time-line}
    \vspace{-0.2in}
\end{figure}

\paragraph*{Terminal condition}
The game terminates at time $T$, when the Mover reaches $g_R$:
\begin{equation*}
    T = \min_{t \in \mathbb{Z}_{\geq 0}} \{t | p(t) = g_R\},
\end{equation*}
which solely depends on the strategy used by the Mover.
Note that the true goal is specified by the nature and does not change throughout the game.
The Eater makes its last action at $t=T-1$ and the banana consumption is updated for one last time (see Fig.~\ref{fig:time-line}). 
The Mover receives the remaining bananas at the true goal.

\paragraph*{Information structure}
The set of goal locations $G=\{g_1,g_2\}$ is a common knowledge.
We consider sequential actions as illustrated in Fig.~\ref{fig:time-line}.
When choosing an action at~$t$, the Mover has access to the states $p(t)$ and $\b(t)$.
The Mover also has private information $I=\{1,2\}$ which indicates the true goal.
The Eater has access to the updated states $p(t+1)$ and $\b(t)$, and  the most recent action $a^{M}(t)$ that the Mover has taken (or equivalently, one-step memory $p(t)$).

\paragraph*{Strategy sets}
We use $\game{i}$, $i\in \{1,2\}$, to denote the game where $g_i=g_R$. 
The strategy of the Mover for $\game{i}$ is  denoted as:
$\pi^M_i:P \times B \times I \rightarrow A_M$,
where $i\in\{1,2\}$.
Note that the dependence on the game parameter, $G$, which is common knowledge, is omitted.
We denote the overall Mover's strategy as 
\vspace{-0.1in}
\begin{equation}
    \pi^M=\begin{cases}
    \pi^M_1&\text{if }g_R=g_1,\\
    \pi^M_2&\text{if }g_R=g_2.
\end{cases}
\end{equation}

Based on the information structure, the Eater's strategy is a mapping
$\pi^E:P \times A_M \times B \rightarrow A_E$.
While the Mover can have different strategy for each game, the Eater must use the same strategy for both games, because it cannot distinguish between them. 

\paragraph*{Objective functions}
\if{0}
The Mover's payoff function for $\game{i}$, $i\in \{1,2\}$ is defined as:
\begin{equation}
    \label{eqn:J-M} J^{M}_i(p,\b;\pi^M_i,\pi^E)=b_i(T(\pi^M_i))
\end{equation}
which the Mover wants to minimize. 

The Eater seeks to maximize this consumption at the true goal. 
However, due to the lack of information on which goal is true, there is a variety of possible preferences for the Eater: e.g., average, worst-case, or best-case performance.
In this paper, we consider the following objective
\begin{equation}
    J^{E}(\pi^M,\pi^E)= \min_{i\in \{1,2\}} J^{M}_i(\pi^M_i,\pi^E)=\min_{i\in \{1,2\}} b_i(T(\pi^M_i)),
    \label{eq:eater_payoff}
\end{equation}
where the dependence on the initial condition is omitted.
This payoff function provides the worst-case guarantee among the two possible games: $\game{1}$ and $\game{2}$.
We will later see that this selection prevents the Eater from being deceived by exaggeration (Lem.~\ref{lem:no_deception}).
\fi

We define the outcome of game $\game{i}$ induced by strategy pair $(\pi^M, \pi^E)$ as
\begin{equation}
    \label{eqn:J} J_i(p,\b;\pi^M_i,\pi^E)=b_i(T(\pi^M_i)),
\end{equation}
which denotes the banana consumed at the true goal when the game terminates.\footnote{The implicit dependency of $b_i$ on $\pi^E$ is omitted.}
Since the Mover knows which game it is playing, we let it directly minimize the consumption at the true goal, and hence its objective function is
\begin{equation}
    J^{M}_i(p,\b;\pi^M_i,\pi^E) = J_i(p,\b;\pi^M_i,\pi^E).
\end{equation}
Although the Eater wants to maximize $J_i$ as well, it cannot distinguish between the two goals. 
Therefore, there is a variety of candidate metrics for the Eater to optimize,
e.g., average, best-case, or worst-case performance. 
In this paper, we set the worst-case performance as the Eater's objective function and let the Eater maximize 
\begin{equation}
    J^E(p,\b;\pi^M_i,\pi^E) = \min_{i\in \{1,2\}} J_i(p,\b;\pi^M_i,\pi^E).
    \label{eq:eater payoff eq4}
\end{equation}

This Eater's objective provides the worst-case guarantee among the two possible games: $\game{1}$ and $\game{2}$. 
We will later see that this objective prevents the Eater from being deceived by exaggeration (Lem.~\ref{lem:no_deception}).
In the sequel, we will omit the dependence of the objective functions on the initial conditions $p$ and $\b$.


We consider the equilibrium concept defined as follows.
\begin{definition}
    \label{def:equilibrium}
    A pair of strategies $(\pi^{M*}, \pi^{E*})$ constitutes an equilibrium, if for all $\pi^E \in \Pi^E$ and $\pi^{M} \in \Pi^{M}$, it satisfies:
    \begin{subequations}
    \begin{align}
         J^{M}_1(\pi^{M*}_1,\pi^{E*})\le J^{M}_1(\pi^{M}_1,\pi^{E*})\label{eq:1} 
\\
               J^{M}_2(\pi^{M*}_2,\pi^{E*})\le J^{M}_2(\pi^{M}_2,\pi^{E*})\label{eq:2} \\
         J^{E}(\pi^{M*},\pi^{E*})\geq J^{E}(\pi^{M*},\pi^{E})\label{eq:3} 
            \end{align}
    \end{subequations}  
    where $\Pi^E$ and $\Pi^M$ are sets of admissible strategies for the Eater and the Mover respectively.
\end{definition}

One may expect the Mover to approach $g_R$ in a ``deceptive'' manner, so that the information regarding the true goal is concealed for as long as possible. 
In the following we propose a pair of strategies and prove that they constitute an equilibrium.

\section{Main results}\label{sec:main}

\subsection{Preliminary Analysis}
We classify the Mover's action based on the change in the distances it induces.
Let $d(\cdot,\cdot): P \times P\rightarrow \mathbb{Z}_{\geq 0}$ denote 1-norm (Manhattan distance) between two points on the grid. 
We use $d_i(t)\triangleq d(p(t),g_i)$, $i\in\{1,2\}$ to denote the distance from the Mover's position $p(t)$ to goal~$g_i$.
We also define the change in the distance as 
\begin{equation}
    \deld{i}(t) \triangleq d_i(t+1) - d_i(t).
    \label{eq: del dist}
\end{equation}

Based on $\deld{1}(t)$ and $\deld{2}(t)$, we categorize the Mover's action at time $t$ into two classes: \emph{\ambiguous{}} and \emph{\revealing{}}.
\begin{definition}[Ambiguous Move]\label{def:ambiguous}
An action of the Mover $a^M(t)$ is an \ambiguous{} move if  $\deld{1}(t) = \deld{2}(t)$.
 \end{definition}
 
 \begin{definition}[Explicit Move]
 \label{def:revealing}
An action of the Mover $a^M(t)$ is an \revealing{} move if 
$\deld{1}(t) = - \deld{2}(t)$.
\end{definition} 

Note that a Mover's action must always fall into either of the above two, since every action results in $|\deld{i}(t)|=1$.

\begin{definition}[Number of Steps]\label{def:number of steps}
For any given $p(t)$, the minimum numbers of \ambiguous{}{} and \revealing{}{} moves required to reach $g_i$ are denoted as $\numa(t)$ and $\numr{i}(t)$.
\end{definition}

    Notice that the number of \ambiguous{} moves cannot be different for the two goals,
    and hence $\numa(t)$ does not have a subscript.
    Also note that $T\geq \numa(0) + \numr{R}(0)$, where the equality holds when the Mover uses a shortest path. 
    Although shortest paths are non-unique, they all contain the same number of \ambiguous{} and \revealing{} moves.
    Finally, the numbers $\numa(t)$ and $\numr{i}(t)$ are independent in the sense that any single Mover action changes only one of them.


We define a quantity that describes the maximum possible banana consumption when the Mover uses a shortest path.
\begin{definition}\label{def: cm}
Let $\cp{i}(t)$ be defined as
\begin{equation}
    \cp{i}(t) \triangleq b_i(t)+d_i(t),
\end{equation}
and $\cm{i}(t)$ be the one measured between the Mover's and Eater's action at time step $t$ (also see Fig.~\ref{fig:time-line}):
\begin{equation}
    \cm{i}(t) \triangleq b_i(t)+d_i(t+1).
    \label{eq: cm}
\end{equation}
\end{definition}
If the Eater had the knowledge of the true goal, it would always eat from that location, but that information is only available to the Mover. Hence one can interpret $\cp{i}(t)$ as the value of the \emph{complete information} game, in which the Eater knows $g_R$ and always eats from the true goal.

\begin{definition}\label{def: diffc}
 Let $\diffcp{i}(t)$ and $\diffcm{i}(t)$ denote the difference functions for $\cp{i}(t)$ and $\cm{i}(t)$ respectively:
 \begin{equation}
 \diffcp{i}(t) \triangleq \cp{i}(t)-\cp{-i}(t)
 \label{eq:diffcp}
 \end{equation}
 \begin{equation}
 \diffcm{i}(t) \triangleq \cm{i}(t)-\cm{-i}(t),
 \label{eq:diffcm}
 \end{equation}
 where $-i=\{1,2\}\setminus\{i\}$.

\end{definition}

\begin{figure}[t]
\centering 
\includegraphics[width=0.34\textwidth]{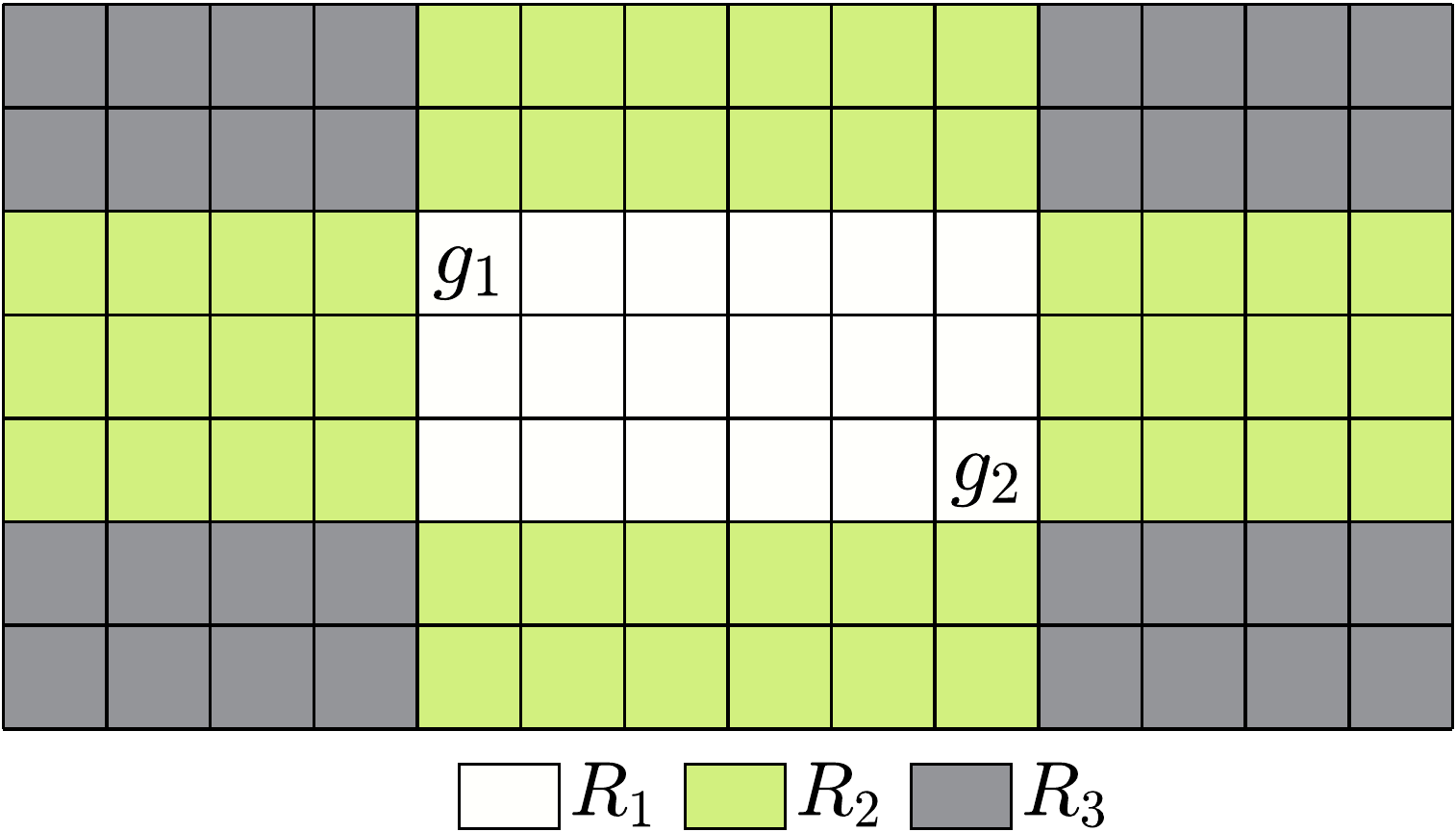}
  \caption{Partition of the game environment.}
  \label{3regions}
  \vspace{-0.2in}
\end{figure}
 
We classify the environment into three regions (See \figref{3regions}). 
$R_1$ is the \emph{no-ambiguity} region. Any move from $R_1$ to $R_1$ is an \revealing{} move. 
$R_2$ is the \emph{partial-ambiguity} region.
Among the two subsets $\{\texttt{right, left}\}$ and $\{\texttt{up, down}\}$, only one will be \ambiguous{} in $R_2$.
$R_3$ is the \emph{full-ambiguity} region, where any action the Mover takes is ambiguous. Note that an action that transfers the Mover from one region to another is \textit{always} ambiguous.

\begin{rem}
Note that for $p(0)\notin R_1$, there always exists a shortest path to $g_i$ that uses all the necessary \ambiguous{} moves first. 
After all the $\numa(0)$ \ambiguous{} moves, the Mover reaches the boundary of $R_1$.
\end{rem}


\subsection{Equilibrium Strategies and Outcome}
We propose the following strategy pair.

\vspace{3pt} \noindent
\textbf{\emph{The Mover's strategy $\pi^{M*}_i$:}} Move on a path with the following two properties. 
\begin{itemize}
\item The path is one of the shortest paths to $g_i$. 
\item All $\numa(0)$ ambiguous moves are made first, and then $\numr{i}(0)$ \revealing{} moves last.
\end{itemize}
\vspace{3pt} \noindent
\textbf{\emph{The Eater's strategy $\pi^{E*}$:}} 
Observe $a^M(t)$ and identify if it was \ambiguous{} or \revealing{} based on Def.~\ref{def:ambiguous} and \ref{def:revealing}.
\begin{itemize}
\item If $a^M(t)$ was \ambiguous{}, then use the \emph{conservative} action:
\begin{equation}
    \vea(t)=\begin{cases}
[1,0],& \text{if\;}\diffcm{1}(t)<0\\
[0,1],& \text{if\;}\diffcm{1}(t)>0\\
[0.5,0.5],& \text{if\;}\diffcm{1}(t)=0.
\end{cases}
\label{eq:ambiguous eater's strategy}
\end{equation}

\item If $a^M(t)$ was \revealing{} then use an \emph{exploiting} action:
\begin{equation}\label{eq:eater's revealing strategy}
    \vea(t)=\begin{cases}
[1,0],& \text{if\;} \deld{1}(t) < \deld{2}(t)\\
[0,1],& \text{if\;} \deld{1}(t) > \deld{2}(t).
\end{cases}
\end{equation}
\end{itemize}


\begin{theorem}[Equilibrium]
\label{thm:equilibrium}
The strategies $\pi^{M*}$ and $\pi^{E*}$ form an equilibrium, i.e., they satisfy equations \eqref{eq:1}, \eqref{eq:2} and \eqref{eq:3}.
\end{theorem}
\begin{proof}
We prove this main result in three steps in Sec.~\ref{sec:proofs}.
Step I: We prove equilibrium for $p(0)\in R_1$ (Lem.~\ref{lem:R1}).
Step II: We prove the optimality of $\pi^{M*}$ outside of $R_1$. 
Specifically, we show that the Mover should use a shortest path (Lem.~\ref{lem:no_move_away} and \ref{lem:no_deception}), and it has no incentive to use \revealing{} moves before \ambiguous{} ones (Lem.~\ref{lem:ambiguous moves first}).
Step III: Optimality of $\pi^{E*}$ outside of $R_1$ is proved (Lem.~\ref{lem: eater conservative strategy}). 
\end{proof}

Before providing the equilibrium outcome, we present some observations that facilitate the analysis throughout the paper.
\begin{rem}\label{rem: delc}
If $a^M(t)$ is \ambiguous{}, then
\begin{equation}\label{eq:diffc ambiguous}
    \diffcm{i}(t)= \diffcp{i}(t).
\end{equation}
Furthermore, if the Eater uses $\pi^{E*}$, then $|\diffcp{i}(t)|$ approaches zero by one after each \ambiguous{} move: i.e.,
    \begin{equation}
    \diffcp{i}(t+1)-\diffcp{i}(t)=\begin{cases}
        1& \diffcp{i}(t)<0\\
        -1& \diffcp{i}(t)>0\\
        0& \diffcp{i}(t)=0.
    \end{cases}
    \label{eq:diff delc}
\end{equation}
\end{rem}
\begin{rem}\label{rem del c revealing}
If $a^M(t)$ is an \revealing{} move toward $g_i$, then
\begin{equation}\label{eq:diffm c revealing}
    \diffcm{i}(t)= \diffcp{i}(t)-2.
\end{equation}
Furthermore, if the Eater uses $\pi^{E*}$, then $\diffcp{i}(t)$ decreases by one: i.e.,
\begin{equation}\label{eq:diffc revealing}
    \diffcp{i}(t+1)= \diffcp{i}(t)-1
\end{equation}
\end{rem}

\begin{theorem}
\label{thm:value}
The equilibrium outcome under $(\pi^{M*},\pi^{E*})$ for game $\game{i}$ is:
\footnote{The time arguments for $p$, $\b$, $\cp{i}$, $\numa$ and $\diffcp{i}$ are omitted for conciseness}
\begin{numcases}{V^{M*}_i(p,\b)=}
   \cp{i}-0.5\numa(1+\mathrm{sgn}(\diffcp{i})) & $\numa \leq|\diffcp{i}|$ \label{eq:16a}
   \\
    \cp{i}-0.5(\numa+\diffcp{i}) & $\numa>|\diffcp{i}|$. \label{eq:16b}
\end{numcases}
The Eater's performance at the equilibrium is then given by
$V^{E*}(p, \b)= \min_{i \in \{1,2\}} V^{M*}_i(p,\b).
    \label{eq: eater's value}$
\label{theorem 2}
\end{theorem}

It is worth noting that the terms $0.5\numa(1+\mathrm{sgn}(\diffcp{i}))$ and $0.5(\numa+\diffcp{i})$ in \eqref{eq:16a} and \eqref{eq:16b} are non-negative, and they represent the reduction in consumption when compared to the complete-information scenario, $\cp{i}(0)$. In this sense, we can interpret these quantities as the \emph{value of information}.

\begin{proof}
If the Mover and the Eater use $\pi^{M*}$ and $\pi^{E*}$ respectively, then $\numr{i}(0)$ \revealing{} moves will be made towards $g_i$, and after each \revealing{} move the Eater will take from $g_i$. This means that $b_i(0)+\numr{i}(0)$ consumption from $g_i$ is guaranteed.
Now we are left with analyzing
the consumption due to $\numa(0)\geq0$ \ambiguous{} moves. 

\emph{Case 1: $\numa(0)\leq|\diffcp{i}(0)|$.} 
From \eqref{eq:diff delc} and $\numa(0)\leq |\diffcp{i}(0)|$ we can conclude that $\diffcp{i}(t)$ will not reach zero. This means for $t\leq \numa(0)$ time steps the Eater will eat only from the goal with lower $\cp{i}(0)$ (see Rem.~\ref{rem: delc}). Then the outcome is
\begin{equation*}
   J^M_i(\pi^{M*}_i,\pi^{E*})=\begin{cases}
        b_i(0)+\numr{i}(0)+\numa(0) &  \text{if\;} \diffcp{i}(0)<0, \\
        b_i(0)+\numr{i}(0) &  \text{if\;} \diffcp{i}(0)>0.
    \end{cases}
\end{equation*}
which can be rewritten as \eqref{eq:16a}.

\emph{Case 2: $\numa(0)>|\diffcp{i}(0)|$.} 
Similar to Case~1, the Eater will keep eating from the goal with lower $\cp{i}$. However, in Case~2, $|\diffcp{i}(t)|$ reaches zero after $|\diffcp{i}(0)|$ steps (see \eqref{eq:diff delc}). 
Then for the remaining $\numa(0)-|\diffcp{i}(0)|$ \ambiguous{} moves, the Eater will take 0.5 from both goals.
Consequently, we have
\begin{align}
    &J^M_i(\pi^{M*}_i,\pi^{E*}) = b_i(0)+\numr{i}(0) \nonumber 
    \\
    &~+\begin{cases}
        |\diffcp{i}(0)|+0.5(\numa(0)-|\diffcp{i}(0)|) &  \text{if\;} \diffcp{i}(0)\leq 0\\
        0.5(\numa(0)-|\diffcp{i}(0)|) &  \text{if\;} \diffcp{i}(0)\geq 0,
        \nonumber 
    \end{cases}
    \label{eq:19}
\end{align}
which can be expressed as \eqref{eq:16b}.
\end{proof}

\section{Proof of Equilibrium}
\label{sec:proofs}
\subsection{Equilibrium in Region $R_1$}

\begin{lem}
\label{lem:R1}
The strategy pair $(\pi^{M*}$,$\pi^{E*})$ forms an equilibrium in $R_1$, i.e., equations \eqref{eq:1}, \eqref{eq:2} and \eqref{eq:3} hold in $R_1$.
\end{lem}
\begin{proof}
Assuming that $p(t) \in R_1$,
we start by analyzing Eater's deviation in $R_1$ (condition \eqref{eq:3}).
From Thm.~\ref{theorem 2}, Eater's outcome in $R_1$ is given by
$J^E(\pi^{M*},\pi^{E*})=\min_{i\in\{1,2\}} \cp{i}(0).$
Consider a different Eater's strategy $\pi^{E'}$, which takes $x\in \{0.5, 1\}$ banana from $g_{-i}$ at least once, even if the Mover is approaching $g_i$.
The Mover's payoff under such strategy is
$J^M_i(\pi^{M*}_i,\pi^{E'}) \leq \cp{i}(t)-x$,
and therefore, we have $J^E(\pi^{M*},\pi^{E'})\leq \min_{i\in\{1,2\}} (\cp{i}(t)-x)=V^{E*} - x$, which implies that the Eater has no incentive to deviate.
Next, we show that the Mover has no incentive to deviate from $\pi^{M*}_i$ in $R_1$.
%
From Thm.~\ref{theorem 2}, the Mover's payoff under $(\pi^{M*}_i, \pi^{E*})$ is given by 
$J^M_i(\pi^{M*}_i,\pi^{E*})=\cp{i}(t)=b_i(t)+\numr{i}(t).$
Consider a different strategy $\pi^{M'}$, which increases the path length by adding additional \revealing{} or \ambiguous{} moves.%
\footnote{Additional \ambiguous{} moves are possible when the Mover is at the boundary of $R_1$.}
In either case, the total number of \revealing{} moves, $\numr{i}(0)$, can only increase but never decrease. 
Therefore, the banana consumption increases, i.e.,  $J^M_i(\pi^{M'}_i,\pi^{E*})\geq b_i(0)+\numr{i}(0) = V^{M*}_i$, 
and thus the Mover has no incentive to deviate.
\end{proof}
\subsection{Mover's Strategy}
This section proves that the Mover has no incentive to deviate from $\pi^{M*}$ when $p(t)\notin R_1$, if the Eater uses $\pi^{E*}$.
\begin{lem}[]
\label{lem:no_move_away}
The Mover has no incentive to deviate from shortest path by moving away from both goals if the Eater uses $\pi^{E*}$.
\end{lem}
\begin{proof}
Consider $\game{i}$ and $p(0)\notin R_1$. 
Under $\pi^{M*}$, the Mover makes
 $\numa(0)>0$ \ambiguous{} moves first and $\numr{i}(0)\geq 0$ later.
Consider $\pi^{M'}\neq\pi^{M*}$, which makes at least one ambiguous move away from both goals.
Clearly, this strategy cannot reduce the minimum number of \revealing{} moves, $\numr{i}(0)$.
%
The question then becomes: by making additional moves away from both goals, can the Mover decrease the resource consumption due to ambiguous moves?

We compare the Mover's trajectory with $\numa(0)$ \ambiguous{} moves and the one with $\numa(0)+2m$ steps, where $m\in\mathbb{N}$ is the number of \ambiguous{} steps \emph{away} from the real goal.
For simplicity, consider the case where $\pi^{M'}$ makes the $\numa(0)$ \ambiguous{} moves first.
Rem.~\ref{rem: delc} implies that in the first $\numa(0)$ steps, $\diffcp{i}(t)$ evolves in the same way for the two trajectories.
Therefore, the consumption in this phase is the same for both trajectories.
The trajectory with $2m$ additional steps has either: (i) no additional consumption if $\diffcp{i}(t)>0$ for $t\leq \numa(0)+2m$, (ii) $2m$ additional consumption if $\diffcp{i}(t)<0$ for $t\leq \numa(0)+2m$, or (iii) somewhere between the above two if $\diffcp{i}(t)=0$ is achieved at some point.
In all these three cases, the consumption at the true goal $g_i$ does not decrease.
This analysis extends to the case where $\pi^{M'}$ does not apply all $\numa(0)$ \ambiguous{} moves first. The detailed proof is omitted due to page constraint.
\end{proof}
One might expect the existence of a deceptive Mover strategy that approaches $g_F$ at the beginning and misleads the Eater to consume more from the fake goal. 
Such behavior is known as an exaggeration~\cite{Topcu2022resal}. The next lemma shows that this is not effective under our problem setting.
\begin{lem}[No Exaggeration]
\label{lem:no_deception}
The Mover has no incentive to make any \revealing{} moves towards $g_F$ if the Eater uses $\pi^{E*}$.
\end{lem}

\begin{proof}
Without loss of generality consider $\game{i}$ where real goal is $g_i$ and fake goal is $g_{-i}$.
We assume $p(0)\notin R_1$ because we already discussed $R_1$ in Lemma \ref{lem:R1}. 
The Mover cannot make any \revealing{} move in $R_3$, so we can restrict our attention to the situation where $p(0)\in R_2$.

Suppose that the Mover uses a different strategy $\pi^{M'}\neq\pi^{M*}$ which makes at least one \revealing{} move towards $g_{-i}$ in $R_2$.
We can see that this ``exaggeration'' move will result in
$\cm{i}(t)=\cp{i}(t)+1$ and
$\cm{-i}(t)=\cp{-i}(t)-1$.
From Def.~\ref{def: diffc} we have
$\diffcm{i}(t)=\diffcp{i}(t)+2$.
{Since $\pi^{E*}$ will respond by taking from $g_{-i}$, we will have $\diffcp{i}(t+1)=\diffcp{i}(t)+1$.}

{
Although the change in $\diffcp{i}(t)$ has no effect on the Eater's behavior after an \revealing{} move,
notice that a larger $\diffcp{i}(t)$ is favorable for the Mover 
based on \eqref{eq:ambiguous eater's strategy}: i.e., the consumption after the \ambiguous{} move.
In this context, the exaggeration move improved $\diffcp{i}(t)$ by 1. 
This increase either: (i) delays a positive $\diffcp{i}(t)$ reaching 0 by one time step, (ii) makes $\diffcp{i}(t)=0$ increase to 1, or (iii) expedites a negative $\diffcp{i}(t)$ reaching zero by 1 time step.
Again, by recalling \eqref{eq:ambiguous eater's strategy}, we can see that this translates to a reduced consumption from $g_i$ after an \ambiguous{} move by at most 1.
}

However, notice that the above ``benefit''  causes an increase in the number of \revealing{} moves remaining. 
Since this always penalizes the Mover by 1 consumption from $g_i$, we can conclude that deception by exaggeration will at best cancel the penalty, but never result in less consumption from $g_i$ in total.
%
\end{proof}

\begin{figure*}[!ht]
    \centering
    \includegraphics[width =0.98\textwidth]{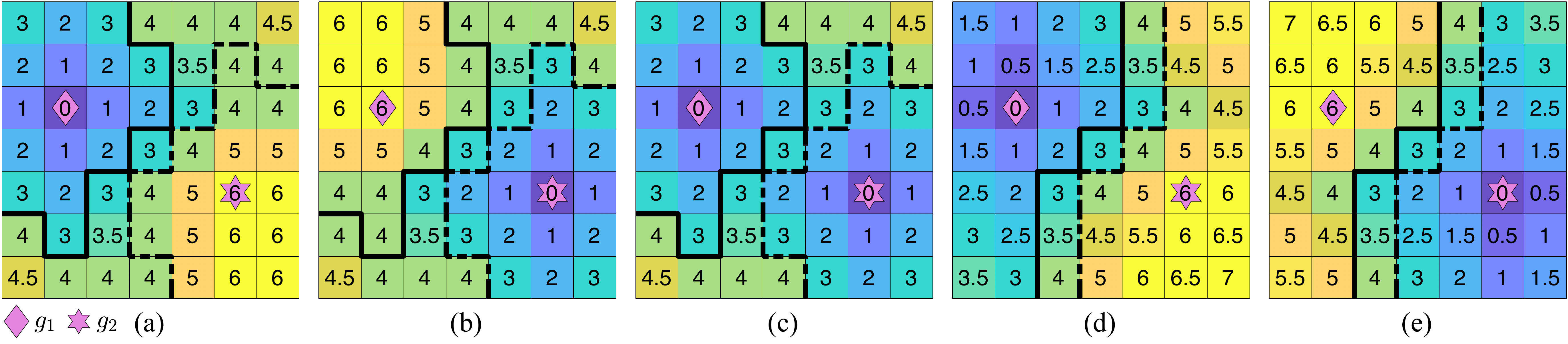}
    \vspace{-5pt}
    \caption{The game outcome under different strategies: (a) $V^{M*}_1$, (b) $V^{M*}_2$, (c) $V^{E*}$; (d)~$J^{M}_1(\pi^{M*}_1,\pi^{E'})$, (e) $J^{M}_2(\pi^{M*}_1,\pi^{E'})$.}
    \label{fig:outcome}
    \vspace{-0.2in}
\end{figure*}

\vspace{-0.1in}
\begin{cor}[Shortest Path]\label{col 1}
The Mover has no incentive to use a non-shortest path if the Eater uses $\pi^{E*}$.
\end{cor}
\begin{proof}
This result directly follows from Lem.~\ref{lem:no_move_away} and~\ref{lem:no_deception}. 
\end{proof}

\begin{lem}[Ambiguous Moves First]\label{lem:ambiguous moves first}
The Mover has no incentive to deviate from making all the ambiguous moves first, given that the Eater uses $\pi^{E*}$.
\end{lem}
\begin{proof}
Without loss of generality suppose the Mover plays $\game{i}$.
From Cor.~\ref{col 1} we know that the Mover must stay on the shortest path towards $g_i$.
The number of steps needed to reach $g_i$ is described as $T = \numa(0) + \numr{i}(0)$.

We would like to compare $b_i(T)$ under the optimal sequence (\ambiguous{} moves first) generated from $\pi^{M*}$,
with the one under $\pi^{M'}\in\Pi^M$, which takes the same number of steps $T(\pi^{M'}) = \numa(0)+\numr{i}(0)$, but in a different order: i.e., at least one \revealing{} move before the final \ambiguous{} move.

Since $\pi^{E*}$ will always take from $g_i$ after observing an \revealing{} move towards $g_i$ regardless of the timing, consumption due to \revealing{} moves under $\pi^{M*}$ and $\pi^{M'}$ will be the same.
Therefore, we focus on the Eater's behavior after the \ambiguous{} moves.
We will prove that making \revealing{} moves earlier will never reduce the subsequent banana consumption from the real goal associated to the \ambiguous{} moves.

Let $t_k^*$ and $t_k'$ for $k\in\{1,...,\numa(0)\}$ denote the times when the $k$-th \ambiguous{} move is used by $\pi^{M*}$ and $\pi^{M'}$, respectively.
Trivially, $t_k^* = k-1$ for all $k$.
The timing for $t_k'$ is delayed by the number of \revealing{} moves used before the $k$-th \ambiguous{} move, which we denote by $\numz\geq 0$.

Now, we will analyze the difference in the banana consumption under the two strategies by looking at $\diffcp{i}(t)$ defined in \eqref{eq:diffcm}.
We will use $\diffcm{i}^*$ and $\diffcm{i}'$ to denote the ones for $\pi^{M*}$ and $\pi^{M'}$, respectively.
Recalling the discussion in Rem.~\ref{rem: delc}, we know that $\diffcm{i}(t)\leq 0$ will result in consumption from the real goal, $g_i$.
Therefore, all we need to show is that
\begin{equation}
    \diffcm{i}^*(t_k^*) \geq \diffcm{i}'(t_k'),\; \forall k\in\{1,...,\numa(0)\}.
    \label{eq:delc_pair_condition}
\end{equation}
The above inequality implies that for the $k$-th \ambiguous{} move, the one from $\pi^{M'}$ will lead to the same or more consumption on the real goal compared to the one from~$\pi^{M*}$.

To see \eqref{eq:delc_pair_condition}, recall Rem.~\ref{rem del c revealing} and see that an \revealing{} move towards $g_i$ at time $t$ will always reduce $\diffcp{i}(t)$ by one.
Also recall that an \ambiguous{} move will make $\diffcp{i}(t)$ approach zero (either from positive or negative side).
Therefore, if $\diffcp{i}(0)\geq 0$, then we have
\begin{align}
    \diffcp{i}^*(t_k^*) &= \max\{0, \;\diffcp{i}(0)-k+1\},\;\text{whereas}\\
    \diffcp{i}'(t_k') &\leq \max\{0, \;\diffcp{i}(0)-k+1-\numz\}.
\end{align}
From Rem.~\ref{rem: delc}, we know that $\diffcm{i}^*(t_k^*) = \diffcp{i}^*(t_k^*)$,
and from Rem.~\ref{rem del c revealing}, we have $\diffcm{i}'(t_k') \leq \diffcp{i}'(t_k')$, where the equality holds when $a^M(t_k')$ is \ambiguous{}.

Thus \eqref{eq:delc_pair_condition} holds for $\diffcp{i}(0)\geq 0$. With a similar argument, we can also show \eqref{eq:delc_pair_condition} for the case when $\diffcp{i}(0)\leq 0$.
\end{proof}

With Cor.~\ref{col 1} and Lem.~\ref{lem:ambiguous moves first}, we have shown that the Mover should stick to $\pi^{M*}$ if the Eater uses $\pi^{E*}$. 

\subsection{Eater's Strategy}
Now we study the optimality of $\pi^{E*}$ when $p(t)\notin R_1$.
Observe that $\pi^{M*}$ uses \revealing{} moves only inside $R_1$, which implies that $\pi^{E*}$ will use the \emph{exploiting} action \eqref{eq:eater's revealing strategy} only in $R_1$.
We therefore focus our attention to the \emph{conservative} action.
\begin{lem}\label{lem: eater conservative strategy}
For $p(t)\notin R_1$, the Eater has no incentive to deviate from its conservative action \eqref{eq:ambiguous eater's strategy}, given that the Mover makes only ambiguous moves according to $\pi^{M*}$.
\end{lem}
\begin{proof}
Consider the effect of Eater's actions outside of $R_1$.
The Eater must consider both $\game{i}$ and $\game{-i}$, but notice that the Mover using $\pi^{M*}$ will make $\numa(0)$ \ambiguous{} moves in both games.
Without loss of generality, we assume $\diffcp{i}(0) \geq 0$ and examine the two cases presented in Thm.~\ref{thm:value}.

\emph{Case 1: $\numa(0) \leq|\diffcp{i}(0)|$.}
Note that $\diffcp{i}(0)$ must be positive in this case.
Under the equilibrium strategies, the Eater will take only from $g_{-i}$ in the first $\numa(0)$ steps, and based on Thm.~\ref{thm:value}, we know the outcome is $V^{E*}=\cp{i}(0)-\diffcp{i}(0)$.
Now, we consider a deviation $\pi^{E'}\neq\pi^{E*}$.
Recalling Rem.~\ref{rem: delc}, we can state $\diffcm{i}(t)>0, \forall t\leq \numa(0)$. The only way for $\pi^{E'}$ to deviate from $\pi^{E*}$ is by eating  $x\in\{0.5,1\}$ from the goal with higher $\tilde{c}(t)$ ($g_i$ in our case) at least once.
Based on \eqref{eq:16a} the Mover's payoff under this strategy will be $J^{M}_i(\pi^{M*}_i,\pi^{E'})\geq \cp{i}(0)-\numa(0)+x$ 
and $J^{M}_{-i}(\pi^{M*}_{-i},\pi^{E'}) \leq \cp{-i}(0)-x = \cp{i}(0)-\diffcp{i}(0)-x$.
Since $\numa(0) \leq|\diffcp{i}(0)|$ the Eater's payoff results in $J^E(\pi^{M*},\pi^{E'})\leq \cp{i}(0)-\diffcp{i}(0)-x<V^{E*}.$

\emph{Case 2: $\numa(0) > |\diffcp{i}(0)|$.}
Notice that $V_i^{M*}=V_{-i}^{M*}$ in Case~2, which is easy to see from \eqref{eq:16b} and the fact that $\cp{i}(t)=\cp{-i}(t)+\diffcp{i}(t)$ and $\diffcp{i}(t)=-\diffcp{-i}(t)$.
Now any deviation $\pi^{E'}\neq\pi^{E*}$ in the first $\numa(0)$ steps will only cause the above equality to break: i.e., $J_i^M(\pi^{M*}_i,\pi^{E'})=V_i^{M*}+0.5x$ and $J_{-i}^M(\pi^{M*}_{-i},\pi^{E'})=V_{-i}^{M*}-0.5x$ for $x\in \mathbb{Z}$.
Such deviation results in $J^E(\pi^{M*},\pi^{E'})=V^{M*}-0.5|x|$, which is suboptimal.
\end{proof}

To summarize, the Eater has no incentive to deviate from its conservative action during Mover's \ambiguous{} moves. This concludes discussion of Eater's and Mover's equilibrium strategies, and provides justification of Theorem~\ref{thm:equilibrium}.

\section{Numerical Illustration}
This section presents the numerical solution to the game environment shown earlier in \figref{fig:game}.
Figures~\ref{fig:outcome}a and \ref{fig:outcome}b show the Mover's performance under $\pi^{M*}$ and $\pi^{E*}$ calculated based on Thm.~\ref{theorem 2}.
We highlight the boundary, where the minimum in \eqref{eq: eater's value}  switches between $\game{1}$ and $\game{2}$. The area between the two surfaces is where $V_1^{M*}=V_2^{M*}$. Figure~\ref{fig:outcome}c shows the Eater's equilibrium performance.

\paragraph*{Suboptimal Eater}
Figures~\ref{fig:outcome}e and \ref{fig:outcome}d show the outcome under an Eater strategy $\pi^{E'}$ that uses exploiting action after \revealing{} moves and $\e=[0.5,0.5]$ after \ambiguous{} moves. 
This strategy may lead to higher banana consumption than $\pi^{E*}$ in certain scenarios. 
For example, in $\game{2}$, if the Mover starts at the top left cell and the Eater applies $\pi^{E'}$, the Eater takes 7 bananas, while under $\pi^{E^*}$ it can only consume 6 according to \figref{fig:outcome}b. 
However, if $\game{1}$ is actually played, $\pi^{E'}$ only takes 1.5 bananas, which is worse than 3 bananas consumed by $\pi^{E*}$. 
Since the Eater does not know which game is played, we let the Eater optimize its worst-case performance as in~\eqref{eq:eater payoff eq4}, and in this sense $\pi^{E*}$ indeed achieves a better worst-case guarantee.

\begin{figure}[h]
   \centering
\includegraphics[width=0.48\textwidth]{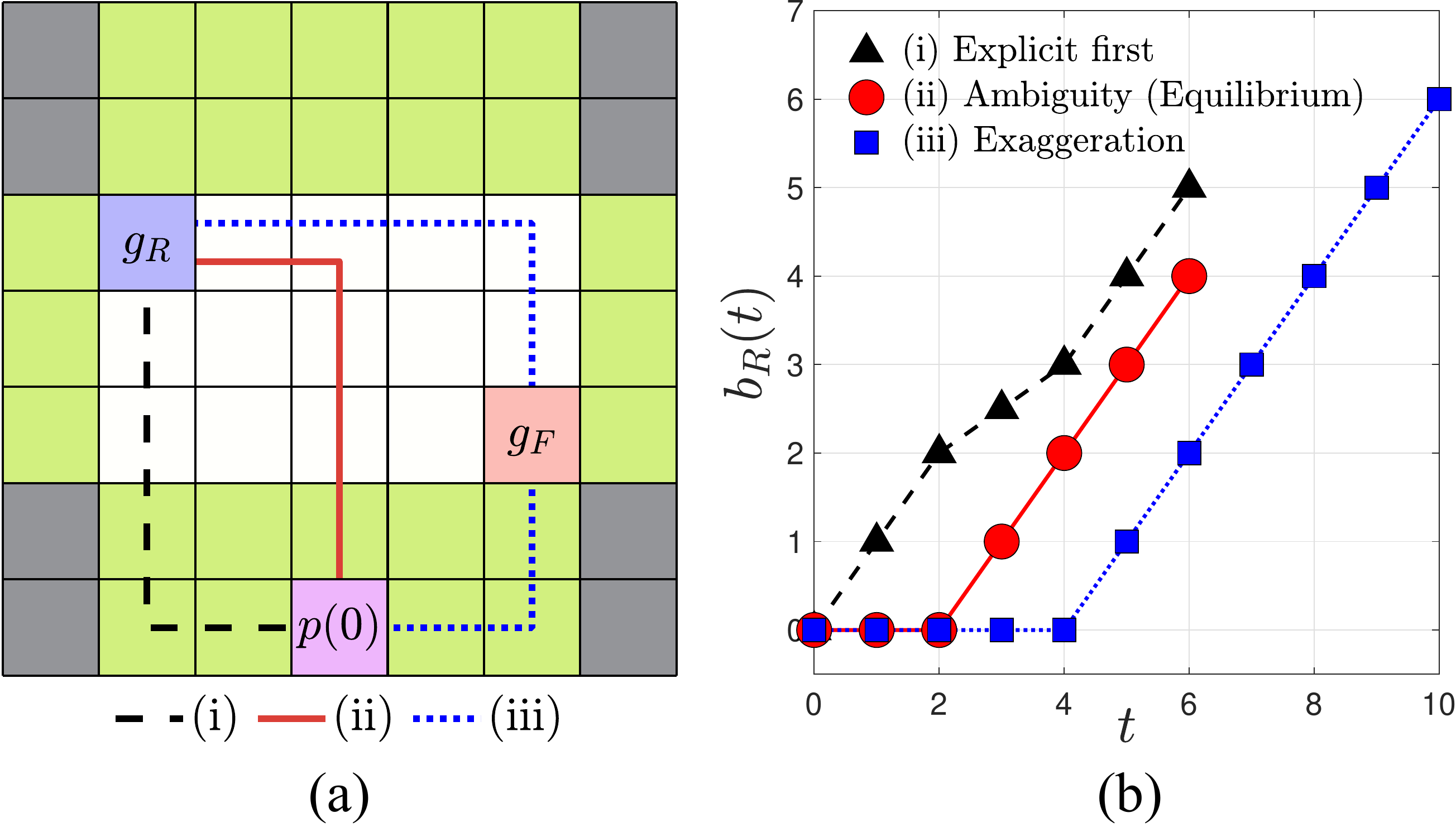}
  \caption{Mover's paths starting with $b_R(0)=b_F(0)=0$. (a) Trajectories in the grid world. (b) Corresponding banana consumption from $g_R$ at each time step under $\pi^{E*}$.}
  \label{3paths}
  \vspace{-0.1in}
\end{figure}
\paragraph*{Suboptimal Mover} 
Figure~\ref{3paths} compares the performance of three different Mover strategies against $\pi^{E*}$.
Path (ii) is the shortest path induced by $\pi^{M*}$, which applies the \ambiguous{} moves first and then the \revealing{} moves. 
At the end of Path (ii), the Eater consumed four bananas (see Fig.~\ref{3paths}b), which is the least amount among the three trajectories, demonstrating the effectiveness of \emph{ambiguity}.
Path (i) is also a shortest path, but the Mover makes \revealing{} moves before the \ambiguous{} ones, which leads to five bananas consumed.
%
Path (iii) corresponds to an exaggeration strategy, which tries to deceive the Eater by moving towards $g_F$ first but actually results in six bananas consumed. 
This is an example of deviating from the shortest path and is clearly suboptimal as discussed in Lem.~\ref{lem:no_deception}. 

\section{Conclusion}
We introduce the Eater and the Mover game, which incorporates an observing agent's action in deceptive motion-control problem. 
We identify a pair of equilibrium strategies, and characterize the equilibrium outcome. 
The results demonstrate that the Mover cannot deceive the Eater by exaggeration if the Eater optimizes its worst-case performance. 
However, the ambiguity is still useful for the Mover to improve its performance. 
Potential extensions of this work include the game with multiple goals and a scenario where the Eater has probabilistic knowledge of the true goal.

 \bibliographystyle{IEEEtran}
 \bibliography{Arxiv_version}

\if{false}
\subsection{Proof of Rem.~\ref{rem: delc}}
\begin{proof}
Under $\pi^{E*}$ the Eater uses conservative action when the Mover makes \ambiguous{} move. To make a decision the Eater compares $\cm{i}(t)$ and $\cm{-i}(t)$. We can look at this difference using \eqref{eq:diffcm}. From \eqref{eq: del dist}, \eqref{eq: cm} and Def.~\ref{def:ambiguous} it follows that:
\[
\cm{i}(t)=b_{i}(t)+d_i(t)+\deld{i}(t)
=\cp{i}(t)+\deld{i}(t)
\] 
\[
\cm{-i}(t)=b_{-i}(t)+d_{-i}(t)+\deld{-i}(t)
 =\cp{-i}(t)+\deld{-i}(t)
 \]
 \[
\diffcm{i}(t)= \cp{i}(t)+\deld{i}(t)-\cp{-i}(t)-\deld{-i}(t) 
 \]
By Def.~\ref{def:ambiguous}: $\deld{i}(t)=\deld{-i}(t)$, then for all \ambiguous{} moves \eqref{eq:diffc ambiguous} holds.

 Since the Eater takes from the a goal with lower $\cp{i}(t)$ factor, we have:
\begin{subequations}
\begin{align}
    \cm{i}(t+1)=\begin{cases}
        b_i(t)+d_i(t+2)+1& \diffcp{i}(t)<0\\
        b_i(t)+d_i(t+2)& \diffcp{i}(t)>0\\
        b_i(t)+d_i(t+2)+0.5& \diffcp{i}(t)=0
    \end{cases}\label{eq: rem3 eq1}\\
    \cm{-i}(t+1)=\begin{cases}
        b_{-i}(t)+d_{-i}(t+2)& \diffcp{i}(t)<0\\
        b_{-i}(t)+d_{-i}(t+2)+1& \diffcp{i}(t)>0\\
        b_{-i}(t)+d_{-i}(t+2)+0.5& \diffcp{i}(t)=0
    \end{cases} \label{eq: rem3 eq2}
    \end{align}
\end{subequations}

 From \eqref{eq:diffc ambiguous}, \eqref{eq: rem3 eq1} and \eqref{eq: rem3 eq2} we obtain \eqref{eq:diff delc}. The equation \eqref{eq:diff delc} means $\diffcp{i}(t)$ approaches zero when $\diffcp{i}(t)\neq 0$, and if it is at zero its value does not change for all following \ambiguous{} moves.
\end{proof}
\fi


 

\end{document}